Storm modulation is feasible through a strategic use of air conditioners

Daisuke Hiruma[1], Ryo Onishi[*2], Keiko Takahashi[2] and Koji Fukagata[1]
[1] Department of Mechanical Engineering, Keio University
[2] Center for Earth Information Science and Technology, Japan Agency for Marine-Earth Science and Technology
*Corresponding Author: onishi.ryo@jamstec.go.jp

(SUMMARY)

Storm trainings, consisting of line-shaped aggregates of cumulonimbi, bring persistent local heavy rains, often causing devastating floods and landslides. Weather control techniques could in theory help prevent such disasters, but so far successful weather control has been limited to local rain initiation or the diffusion of local clouds. No successful strategies have been proposed for the control of mesoscale storms. Here we show that a strategic use of consumer air conditioners, which can typically remove about 1kg of moisture from the air per hour when run in dehumidification mode, and which are installed in large numbers in big cities, can modulate a storm downstream. We numerically reproduced a storm training that affected the Hiroshima metropolis in Japan in 2014, and conducted experiments to test the sensitivity of the storm to the initial moisture field near the surface. We propose an empirically-derived formula for a control efficiency parameter, which can be used to estimate the impact of moisture removal on the rainfall accumulation. It reveals that removal of half a kiloton of moisture, which could be achieved within half an hour in a city with a population of one million since more than one air conditioner is installed per capita in Japan, could lead to a significant reduction of the total rainfall accumulation over a 100 km$^2$ area of heavy rain during the storm event. Conversely, our results indicate that some summertime storms occurring inside or near a metropolis could be strengthened by the excess moisture discharged from large numbers of air conditioners used for cooling rooms. We anticipate our results, which reveal that human activity can have a significant impact on storms, will be a starting point for considering the coupling of weather and the economy, and will contribute to the development of a sustainable society.

(TEXT)

Line-shaped rain bands - so-called storm trainings[1-4] - are often observed in heavy rainfall events, which often cause serious damage to society. Kato (2015)[5] defined a storm training as a line-shaped aggregate of cumulonimbi with a width between 20 and 50 km and a length between 50 and 200 km. Such systems often persist over a given location for a few hours, bringing extremely high rainfall accumulations. In Japan, two-thirds of heavy rain events excluding those related to typhoons can be attributed to storm trainings[6]. For example, it is reported that the serious heavy rain event in the west of Japan in July 2018 was brought about by several tens of storm trainings[7].

Owing to recent advancements in observational and computational technologies, detailed investigation and forecasting of storm trainings has now become feasible, and with these advances comes the possibility of modifying storm trainings to prevent or reduce disasters. The field of weather modification has a long history. The main focus so far has been on methods to artificially modify rainfall. Langmuir (1961)[8] found that the introduction of dry ice and iodide into a sufficiently moist cloud could induce precipitation. These substances work as cloud condensation nuclei (CCN) or ice-forming nuclei (IN), which promote the generation of cloud droplets or ice crystals, and also the conversion of super-cooled liquid droplets into ice crystals. This kind of dispersion of CCN or IN into a cloud is called cloud seeding, and is mainly used to alleviate droughts. Seeding techniques have been developed by many researchers working in many countries[9-11], and about 100 experiments are conducted every year. Recently, hurricane control was also discussed. Hoffman (2004)[13] numerically simulated two destructive hurricanes, Iniki and Andrew, which struck in 1992. The results showed the possibility of changing the track and weakening the hurricanes by making relatively small changes to the initial sea surface temperatures. Rosenfeld et al. (2007)[14] investigated that the possible impact of seeding with submicron CCN on hurricane structure and intensity. However, a realistic and feasible method for controlling hurricanes is yet to be identified.

Cloud seeding modifies the inner structure of individual clouds, whose horizontal length scales are typically around 10 km. Hurricane control would require modification of mesoscale cloud systems with typical length scales of around 1,000 km. No attempts have so far been made to modify storm training systems, whose length scales are typically around 100 km.

Here we show the feasibility of modifying storm training systems by making use of consumer air conditioners installed in metropolitan areas. According to national statistics, more than one consumer air conditioner is installed per capita in Japan (Japanese national census 2015; Japanese cabinet office consumption trend survey 2019). If all these air conditioners were operated together in a dehumidification mode, very large amounts of water vapor (moisture) could be removed from the lower atmosphere. We show that such levels of moisture removal can have a significant impact on a storm training.

**STORM TRAINING SYSTEM**

We conducted numerical weather simulations for a western Japan domain (Fig. 1a) using a model with a 500 m horizontal grid spacing for the storm training event that occurred over Hiroshima on 19$^{th}$ and 20$^{th}$ August 2014, which led to devastating landslides. The reference simulation (Fig. 1b) successfully produces a line-shaped heavy rainfall area, with a length of approximately 70 km in the south-west to north-east direction and a width of approximately 20 km. The most intense simulated rainfall area is located on the east side of Hiroshima city, and is shifted eastward by only around 20 km compared with JMA radar observations (Extended Data Fig. 1a). The typical three-dimensional shape of a storm training, consisted of a chain of cumulonimbus lines, is also well reproduced (Fig. 1c).

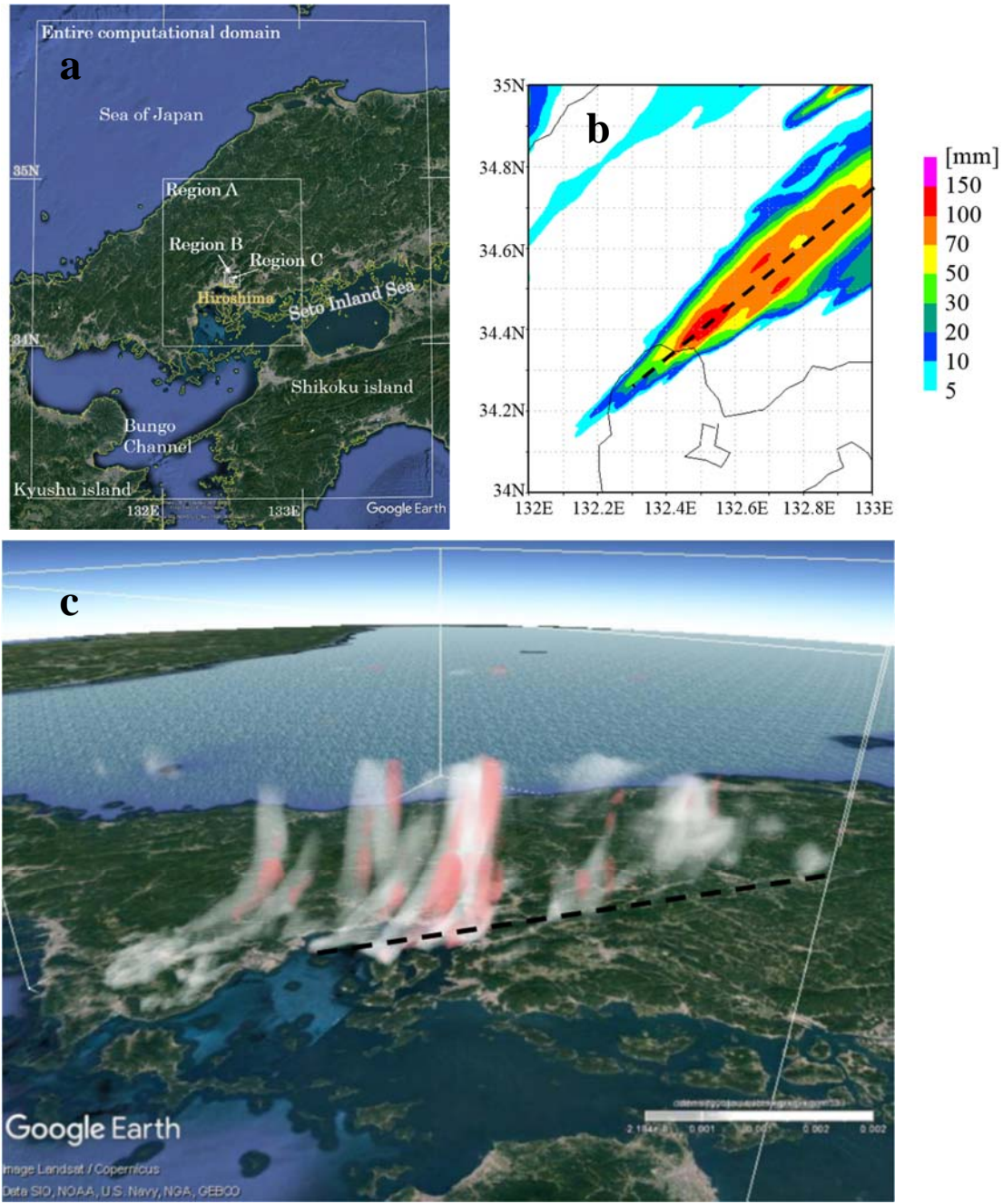

**Fig. 1 | Numerical weather simulation. a**. The computational and analysis domains. **b.** The horizontal distribution of the 6-hour (11 pm - 5 am) rainfall accumulation over Region A. **c**. A three-dimensional visualization of the cloud distribution over Region A at 00:10 am viewed from the southeast direction, visualized using the open software named VDVGE[15]. For this visualization, we ran a 100 m grid-spacing simulation for 10 minutes starting at 12 am from a linearly-interpolated 500 m grid-spacing model state. The total mixing ratios of non-sedimenting hydrometeors - i.e. cloud water and cloud ice - are shaded in white, whereas those of sedimenting hydrometeors - i.e., rain,

graupel and snow – are shaded in red. The images are stretched by a factor of three in the vertical direction to aid recognition of the convective system. The dashed line in **c** corresponds to that in **b**.

**Moisture removal and control efficiency**

We conducted sensitivity experiments in which water vapor mixing ratios below 500 m were reduced upstream of the storm training. We mostly changed the moisture over the Bungo Channel, where the lower atmosphere contained large moisture concentrations (Fig. 2), since a persistent south-easterly wind blew into Hiroshima during the storm. We reduced moisture levels below 500 m at the same uniform rate in all grid boxes over (i) the entire 8,660 km$^2$ computational domain (Case EN), (ii) sub-domains with areas of 4,800 km$^2$ (cases PA-1 and PA-2), and (iii) sub-domains with areas of 100 km$^2$ (cases PB1-9) (Fig. 2). We analyzed the impact of the moisture removals on rainfall over several rectangular target regions – specifically, the rainfall accumulations for the 6-hour period from 11pm on 19th to 5am on 20th August 2014. Four target regions with different sizes were chosen (Fig.1a): (1) the entire 8,660 km$^2$ computational domain (Region EN), (2) region A, which has an area of 1,000 km$^2$, (3) region B, which has an area of 100 km$^2$, and (4) region C, which has an area of 10 km$^2$. Regions B and C are focused on the largest rainfall accumulation area in Hiroshima. The storm training structures were not disrupted by the moisture removals, but the intensity of the rainfall was modulated (see an example in Extended Data Fig. 2).

In order to quantify the performance of the system, we define a control efficiency $E_f$ as

$$E_f = O_b / M_r, \qquad (1)$$

where $O_b$ is the total amount of the rainfall reduction over the target domain for the 6 hours from 11 pm to 5 am, and $M_r$ is the amount of moisture removal in the 500 m high control volume.

Fig. 3a shows the control efficiency for Region A as a function of the amount of moisture removal $M_r$. The efficiency is larger for smaller amounts of moisture removal. For each control area we get slopes of around -0.5. Fig. 3b shows the control efficiency normalized by $M_r^{-0.5}$ as a function of the ratio between the target and control areas $R_S$ =$S_{trg}$ /$S_{ctr}$, where $S_{trg}$ is the area of the target region and $S_{ctr}$ the area of the control region. This shows a collapsed envelope, which corresponds to the maximum attainable efficiency.

The maximum impact of moisture removal can be estimated from the envelope curve. For example, the curve predicts an $E_f$ /$M_r^{-0.5}$ value of $10^6$ for $R_S$=1 (solid star in Fig.3b). For the case of a target region of area 100 km² and a control area of 100 km² (i.e., $R_S$=1), 500 tons of moisture removal would lead to $E_f$=1,400, which would reduce the rainfall accumulation by 7.0 mm over the 100 km² target region. For the case of a target region with area 10 km² and a control area of 100 km² - that is, for $R_S$=0.1 - $E_f$ /$M_r^{-0.5}$ would be $4\times10^5$ (open star in Fig.3b), so the same amount of moisture removal would reduce the rainfall accumulation by 28.0 mm depth over the 10 km² target region. More than one consumer air conditioner is installed per capita in Japan. Typically, such air conditioners can remove about 1 kg of water per hour from the air when they are run in dehumidification mode. Thus, if all the consumer air conditioners in a city with a population of one-million people – similar to the population of Hiroshima – were to operate together in dehumidification mode, around 500 metric tons of water would be removed from atmosphere every 30 minutes. The 7.0 mm accumulated rainfall reduction over a 100 km² region corresponds to nearly 10% of the accumulated rainfall amount of 77.9 mm over Region B, and the 28.0 mm reduction over 10 km² to 20 % of the 132.7 mm rainfall accumulation over Region C. We can thus conclude that a feasible modification of the moisture field can be used to significantly modulate local rainfall accumulations during a storm training event if reliable numerical forecasts are available.

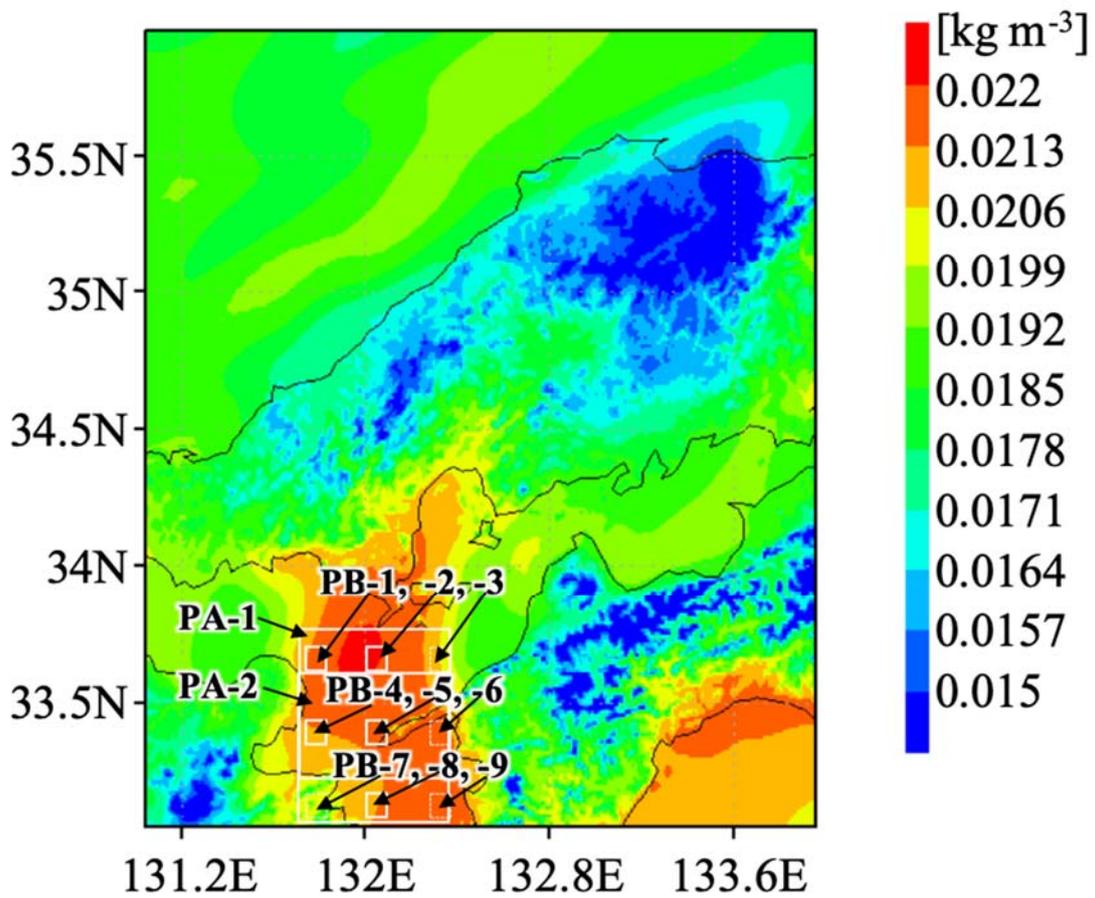

**Fig. 2 | Horizontal distributions of initial water vapor at 300 m over the entire computational domain.** The rectangles show control regions with areas of $10^4$ km$^2$ (areas PA-1 and PA-2) and 100 km$^2$ (areas PB-1 to -9). The control efficiencies for Region A obtaining by controlling the moisture over areas PB-3, -6, -7 and -9, indicated by dashed rectangles, were found to be negative.

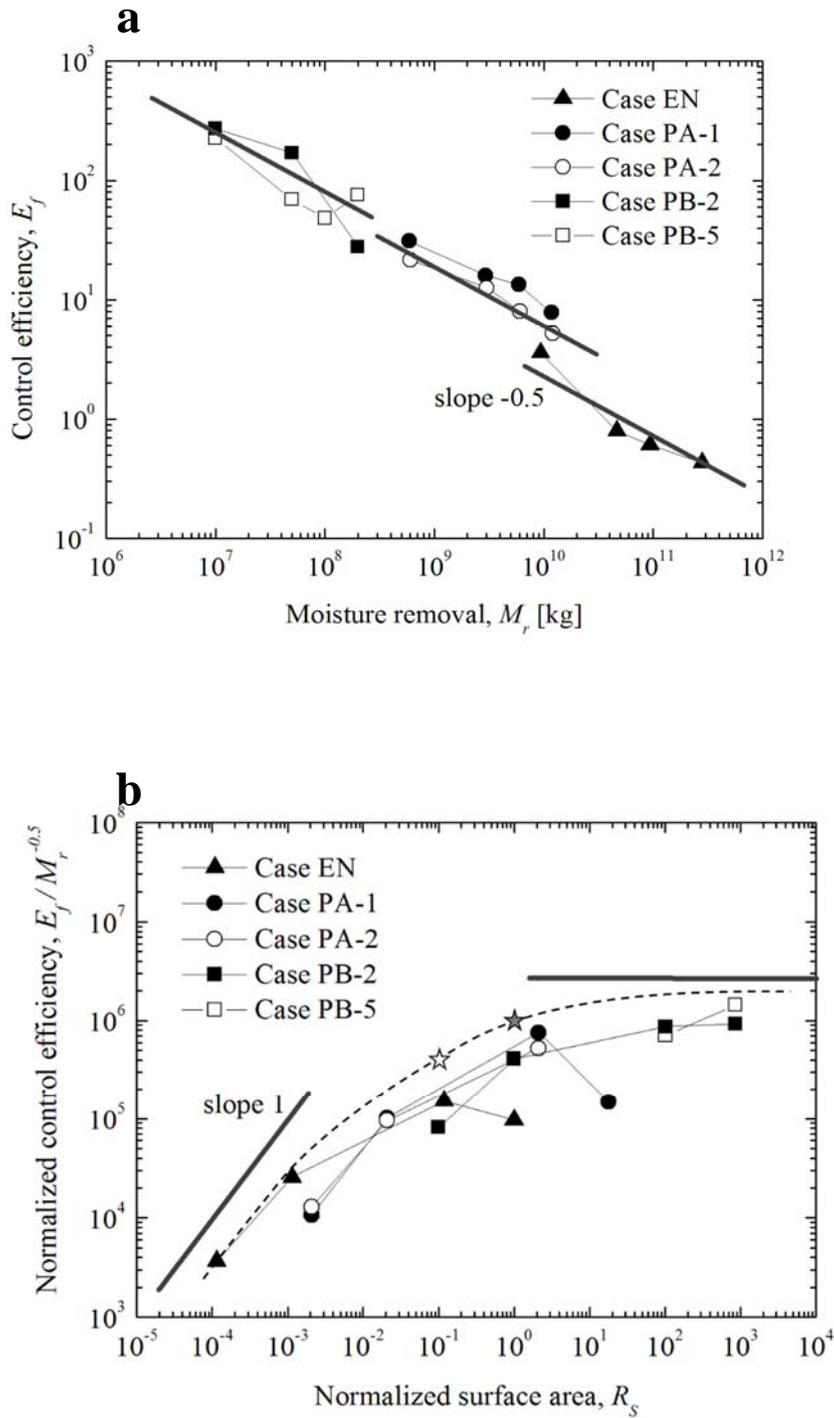

**Fig. 3 | Control efficiency. a**. Control efficiency for Region A as a function of moisture removal. **b.** Normalized control efficiency as a function of the normalized area, $R_S = S_{trg}/S_{ctr}$. The dashed line shows the expected maximum attainable efficiency, corresponding to the upper-bound envelope of the calculated values. The maximum attainable

efficiency increases in proportion to $R_S$ for $R_S \ll 1$, and saturates for $R_S \gg 1$. These tendencies can be explained by the illustration in Extended Data Fig. 3. The values are listed in Extended Data Table 1.

**CONCLUSION**

We have demonstrated the feasibility of modulating a storm training that appeared over Hiroshima on 19th and 20th August 2014 by conducting sensitivity test simulations with reduced moisture in the lower atmosphere. We confirmed that a small upstream removal of moisture can be magnified via a nonlinear response to give a reduction of local rainfall accumulations that can be greater by a factor of one thousand. We proposed empirical calculation of a curve that indicates the maximum attainable control efficiency against the ratio of the target area to the control area. Estimations made with the curve calculated for the case under consideration confirm the feasibility of modulating the rainfall accumulation due to the storm training if moisture near the surface can be controlled through use of the numerous consumer air conditioners that are installed in megacities. Conversely, this also implies that the massive use of air conditioners for cooling rooms in summertime can sometimes strengthen storm training events.

**Acknowledgements**

The numerical simulations were conducted on the Earth Simulator of the Japan Agency for Marine-Earth Science and Technology.

**Author contributions**

D.H. performed numerical simulations and analyzed the results. R.O. designed the work and prepared the manuscript. K.T. supervised the work. K.F. supervised the analysis.

**Competing interests**

The authors declare no competing interests.

# METHODS

**Numerical Weather Simulations.**

The numerical simulations were carried out using the Multi-Scale Simulator for the Geoenvironment (MSSG) [16-18]. MSSG is an atmosphere-ocean coupled non-hydrostatic model aimed at seamless simulations from global to local scales. In this study, it was configured as a meso-scale atmospheric simulation model. The governing equations for the dynamics are the transport equations for compressible flow, which consist of the conservation equations for mass, momentum, energy, and water vapor. The third-order Runge-Kutta scheme was used for time integrations. A fifth-order upwind scheme is chosen for momentum advection and the second-order weighted average flux scheme with the Superbee flux limiter[19] was selected for scalar advection. For turbulent diffusion, the Mellor-Yamada-Nakanishi-Niino level 2.5 scheme[20] was used. The MSSG-Bulk model[16], which is a six-category bulk cloud microphysics model, was used for explicit cloud physics. The Model Simulation radiation TRaNsfer code (MSTRAN)[21] was used for calculating longwave and shortwave radiation transfer.

The horizontal computational domain includes a part of the Chugoku, Shikoku, and Kyushu regions (Fig. 1a). The horizontal grid spacing $\mathit{\Delta x_H}$ was set to 500 m. The domain height was 30 km, with the vertical grid spacing $\mathit{\Delta x_V}$ varying from 14 m for the lowest layer to 123 m for the highest layer. The number of grid points was $N_\lambda \times N_\varphi \times N_z$ =648 × 648 × 207. The simulation start time was set to 9 pm on 19th August 2014 and the end time to 5 am the next day. For the initial and lateral boundary conditions, we utilized the JMA-Meso Analysis (MANAL), which is provided by the Japan Meteorological Agency (JMA). The MANAL dataset has a horizontal grid spacing of 5 km and a temporal resolution of 3 hours. The dataset was interpolated linearly in time and space in order to provide high-resolution input data for MSSG.

The rainfall accumulation map calculated from the JMA radar observations shows a similar pattern to the simulated pattern, with large accumulations extending in a line from the south-west to the north-east (Extended Data Fig.1). The simulated rainfall line shown in Fig.1b is shifted slightly eastward from the observed one by only around 20 km. This comparison confirms the successful reproduction of the storm training in the present MSSG simulation.

**Data availability**

The data is available from the corresponding author upon reasonable request. The numerical simulation code is available under a collaborative framework with the corresponding author.

# Extended Data Figures

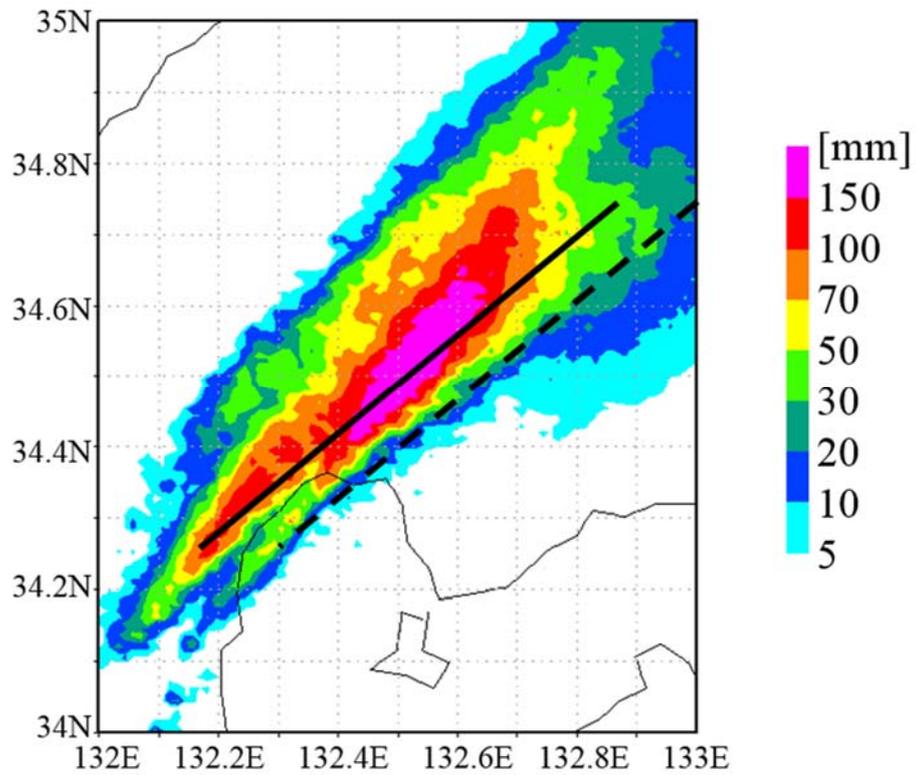

**Extended Data Fig.1 | Observed rainfall accumulation.** Horizontal distribution of the rainfall accumulation over Region A for 6 hours from 11 pm to 5 am, as calculated from JMA radar observations. The dashed line corresponds to that in Fig.1b.

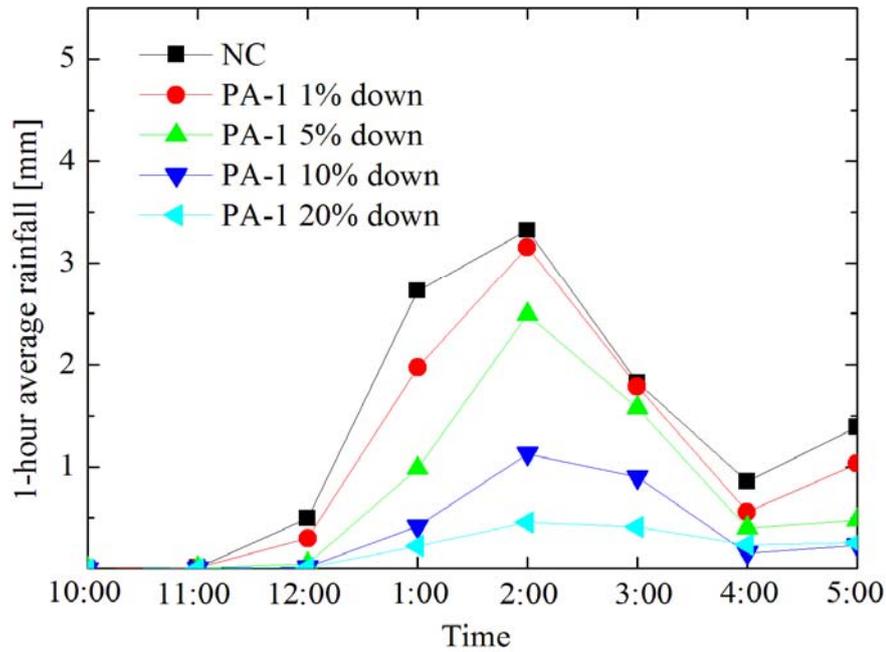

**Extended Data Fig. 2 | Time series of domain-averaged values of 1-hour rainfall accumulation in Region A for case PA-1.** The water vapor mixing ratios over area PA-1 below 500 m height were reduced by the percentages denoted in the legend. Larger reductions in moisture lead to larger reductions in rainfall, but the response is not linear, as clearly shown by the control efficiencies plotted in Fig. 3a.

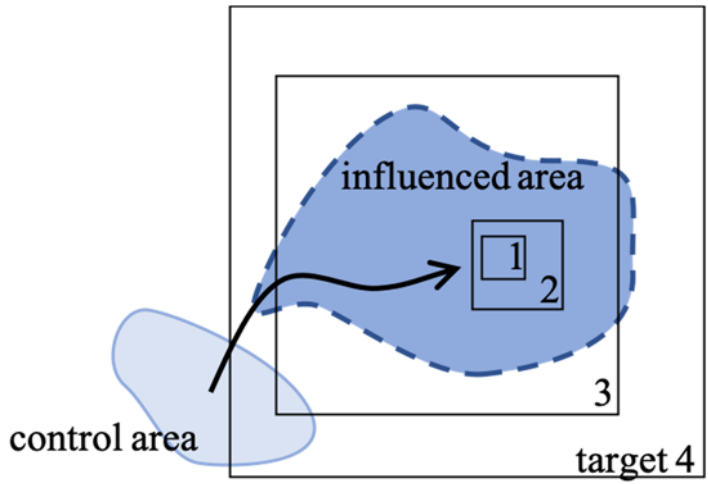

**Extended Data Fig. 3 | Schematic illustration of the influence of moisture removal.** The influence can be magnified through non-linear cloud dynamics, and the influenced area is advected by the wind and expanded by propagation and diffusion. For small target regions, i.e., $S_{trg} \ll S_{ctl}$, the total amount of rainfall reduction is simply proportional to $S_{trg}$ (compare regions 1 and 2). That is, the control efficiency is proportional to the ratio of the target and control areas $R_S$ ($=S_{trg}/S_{ctr}$). For $S_{trg} \gg S_{ctl}$ - i.e., when the target region is much larger than the influenced area - the total amount of rainfall reduction becomes insensitive to $S_{trg}$ (compare regions 3 and 4), indicating a saturation of the control efficiency.

|          | Moisture reduction $M_r$ in kg | Entire domain | Region A | Region B | Region C |
|----------|-------------------------------|---------------|----------|----------|----------|
| Case EN   | $9.37\times10^9$  | 1.00     | 1.56   | 0.27     | 0.04     |
| Case PA-1 | $5.89\times10^8$  | 6.09     | 31.14  | 4.22     | 0.44     |
| Case PA-2 | $6.03\times10^8$  | (-0.93)  | 21.49  | 3.89     | 0.53     |
| Case PB-1 | $9.95\times10^6$  | (-16.2)  | 104.75 | (-5.84)  | (-2.13)  |
| Case PB-2 | $1.02\times10^7$  | 290.58   | 272.28 | 128.56   | 25.63    |
| Case PB-4 | $9.72\times10^6$  | (-83.75) | 13.84  | (-9.57)  | (-3.58)  |
| Case PB-5 | $1.00\times10^7$  | 457.08   | 224.99 | (-15.92) | (-6.75)  |
| Case PB-7 | $9.95\times10^6$  | (-16.17) | 104.75 | (-5.84)  | (-2.13)  |

**Extended Data Table 1 | Control efficiencies for different target regions for the 1% moisture reduction cases.** The cases with positive values for region A are listed. Negative values, signifying increases in rainfall, were obtained more for smaller controlled areas.